\documentclass[aps,prd,superscriptaddress,twocolumn]{revtex4}

\usepackage{graphicx}

\begin{document}

\title{Kinematics of discretely self-similar spherically symmetric
spacetimes} 

\author{Carsten Gundlach}
\email[]{C.Gundlach@maths.soton.ac.uk} 
\affiliation{Faculty of Mathematical Studies, University of Southampton,
         Southampton SO17 1BJ, UK}

\author{Jos\'e M. Mart\'\i n-Garc\'\i a}
\email[]{J.M.Martin-Garcia@maths.soton.ac.uk} 
\affiliation{Faculty of Mathematical Studies, University of Southampton,
         Southampton SO17 1BJ, UK}
\affiliation{Dep.\@ Matem\'atica Aplicada y M\'etodos Inform\'aticos,
ETSI Minas, UPM, Madrid 28003, Spain}

\date{31 May 2003}


\begin{abstract}

We summarize the consequences of the twin assumptions of (discrete)
self-similarity and spherical symmetry for the global structure of a
spacetime. All such spacetimes can be constructed from two building
blocks, the ``fan'' and ``splash''. Each building block contains one
radial null geodesic that is invariant under the self-similarity
(self-similarity horizon).

\end{abstract}


\maketitle



\section{Introduction}


Continuous self-similarity (CSS), or homothety, in a spacetime
expresses the absence of a preferred length or timescale. As a
continuous symmetry, it reduces the number of coordinates on which the
metric depends by one. In combination with spherical symmetry, it
therefore reduces the Einstein equations to a system of ordinary
differential equations, and this is one reason why spacetimes with
these combined symmetries have been studied extensively. Besides
simplicity, a general reason for studying solutions with symmetries in
physics is that attractors of the time evolution flow in phase space
usually have more symmetries than generic solutions. For example, in
general relativity the end state of gravitational collapse is a
Kerr-Newman black hole, which is stationary and axisymmetric.

From a physical point of view, spherically symmetric CSS solutions
were studied, among other reasons, because they give rise to naked
singularities from regular initial states
\cite{OriPiran,OriPiran2}. It has been suggested that the formation of
a naked singularity is always associated with self-similar collapse
\cite{Lake}.  These singularities did not seem, however, to be
attractors of the evolution flow (but see \cite{HaradaMaeda}).

Then Choptuik \cite{Choptuik} found a self-similar solution acting as
an {\it intermediate attractor} in gravitational collapse close to the
black hole formation threshold. The effects seen by Choptuik with
scalar field matter were confirmed by Abrahams and Evans
\cite{AbrahamsEvans} in the collapse of axisymmetric gravitational
waves and by Evans and Coleman \cite{EvansColeman} in spherical fluid
collapse with the scale-free equation of state $p=\rho/3$. For a
review see \cite{critreview}. 

In all these cases the intermediate attractor, or critical solution,
is self-similar and contains a naked singularity. The fluid critical
solution is CSS, but the scalar field critical solution and the
gravitational wave critical solution display a discrete
self-similarity (DSS). Such a symmetry had not previously been
considered in general relativity (but see \cite{Sornette} for DSS in
other sciences).

CSS spherical perfect fluid solutions were generated numerically by
Ori and Piran \cite{OriPiran2}. A family of CSS spherical scalar field
solutions was obtained in closed form by Roberts \cite{Roberts}. All
such solutions were later classified by Brady \cite{Brady}. A
semi-kinematical classification of CSS spherical null dust spacetimes
was carried out by Nolan \cite{Nolan}.

Here, we review the purely kinematical consequences (that is, not
using the Einstein equations) of self-similarity combined with
spherical symmetry. Our methods are generalizations of those in
\cite{OriPiran2,Nolan}. The results of this paper have already been
applied to a classification of all CSS spherical perfect fluid
solutions \cite{fluidcss}, and to the scalar field critical solution,
which is DSS \cite{Choptuikglobal}.


\section{Geometric self-similarity}



\subsection{Definition and adapted coordinates}


To put our results for spherical spacetimes into a wider context, we
first discuss self-similarity without assuming any other symmetry. In
particular, we do not yet restrict to spherical symmetry. A spacetime
is continuously self-similar (CSS), or homothetic, if there is a
vector field $\xi^a$ such that
\begin{equation}
{\cal L}_\xi g_{ab} = -2 g_{ab}.
\end{equation}
The value $-2$ on the right-hand side is a convention that normalizes
$\xi^\mu$, but it must be constant. Any additional structure that
covariantly defines an orientable vector field (for example, the
4-velocity vector of a fluid) allows to generalize the concept of
continuous self-similarity beyond homothety \cite{CarrColey}, but we
will not consider such generalizations here.

In a CSS solution the integral curves of the homothetic vector field
(CSS lines) are geometric objects, and hence provide a geometrically
preferred fibration of the spacetime. We choose coordinates such that
they are lines of constant coordinates $x^i$, $i=1,2,3$. The metric
can then be written in the form
\begin{equation}
\label{generalDSS}
g_{\mu\nu}=e^{-2\tau} \bar g_{\mu\nu},
\end{equation}
where $x^\mu=\{x^0\equiv \tau,x^i\}$, and where $\bar g_{\mu\nu}$
depends on the $x^i$ but not on $\tau$.

A spacetime is discretely self-similar (DSS) if there is a conformal
isometry $\Phi$ of the spacetime such that
\begin{equation}
\label{DSSgeometric}
\Phi_* g_{ab}=e^{-2\Delta}g_{ab}. 
\end{equation}
$\Delta$, the dimensionless scale period, is made unique by taking it
to be the smallest positive value for which (\ref{DSSgeometric})
holds. The value of $\Delta$ is then a geometric property of the
spacetime.  The metric of a DSS spacetime can also be written in the
form (\ref{generalDSS}), where $\bar g_{\mu\nu}$ is now allowed to
depend periodically on $\tau$ with period $\Delta$. 

The form (\ref{generalDSS}) of the DSS metric is preserved under
coordinates changes of the form
\begin{eqnarray}
\label{dssxtransform}
{x'}^i &=& \varphi^i(\tau,x^j), \\
\tau' &=& \tau + \psi(\tau,x^j),
\end{eqnarray}
where $\varphi^i$ and $\psi$ are periodic in $\tau$ with period
$\Delta$. In analogy with CSS lines we call the lines of constant
$x^i$ DSS lines, but unlike the CSS lines they are not unique. The
non-uniqueness is parameterized by (\ref{dssxtransform}).

A closely related type of coordinate system frequently used for CSS or
DSS spacetimes is obtained when one replaces $\tau$ by $t\equiv
-e^{-\tau}$. (The two minus signs are merely a matter of
convention. Note that $t$ is not necessarily a time coordinate.)  The
metric becomes
\begin{equation}
\label{CSS_coordinates2}
ds^2=\bar g_{00}\, dt^2 - 2t\,\bar g_{0i} \,dt\,dx^i + t^2\, \bar
g_{ij}\,dx^i\,dx^j ,
\end{equation}
where the metric coefficients $\bar g_{\mu\nu}$ are the same as in
(\ref{generalDSS}). 
If one also replaces the $x^i$ by $r^i\equiv (-t)x^i$, the
metric becomes
\begin{eqnarray}
\label{CSS_coordinates3}
\nonumber
ds^2&=&(\bar g_{00}+2\bar g_{0i}x^i+g_{ij}x^ix^j)\, dt^2 \\
\nonumber
&+& 2(\bar
g_{0i}+\bar g_{ij}x^j)dt \,dr^i +\bar g_{ij}\,dr^i\,dr^j \\
&\equiv & \gamma_{00}\, dt^2 + 2\gamma_{0i}\,dt\,dr^i
+\gamma_{ij}\,dr^i\,dr^j.
\end{eqnarray}
Note that all coordinates can be spacelike, null, or timelike, as we
have not made any assumptions about the sign of the metric
coefficients. Note also that in coordinates $t$ and $r^i$ the metric
coefficients $\gamma_{\mu\nu}$ still depend only on $x^i= r^i/(-t)$ in
the CSS case, and in the DSS case also periodically on
$\tau=-\log(-t)$.

It is natural to think of the coordinates $x^i$ and $\tau$ as
dimensionless, and of $t$ and $r^i$ as having dimension length.  In
this case one should define $t\equiv -l_0e^{-\tau}$, where $l_0$ is a
fixed arbitrary length scale, and similarly in other expressions. In
the following we do not write $l_0$ for simplicity.


\subsection{Geometric structure}


The decomposition (\ref{generalDSS}) of the metric suggests that we
classify elements of the spacetime as ``kinematical'' or
``dynamical'' depending on whether they are generated by the conformal
factor $e^{-2\tau}$ or the structure of $\bar{g}_{\mu\nu}$. Some
of the effects of the conformal factor can be understood by
dimensional analysis. For example, the Kretschmann scalar $K$, which
has dimension (length)${}^{-4}$, scales as $e^{4\tau}$ along CSS or
DSS lines:
\begin{equation}
K(\tau,x) = e^{4\tau}\bar{K}(\tau,x),
\end{equation}
where $\bar{K}$ depends on $x$, and in DSS also periodically on
$\tau$. ($\bar{K}$ is not the Kretschmann scalar of
$\bar{g}_{\mu\nu}$.) All other curvature scalars scale similarly,
according to their dimension. Therefore $\tau=\infty$, for any $x^i$,
is {\it generically} a scalar curvature singularity of any self-similar
spacetime. We call it the ``kinematical singularity''. However,
$\bar{K}=0$ or $\infty$ can occur on an isolated CSS or DSS line, say
$x=0$, and in this case the simultaneous limit $x\to 0$,
$\tau\to\pm\infty$ may depend on the curve $x(\tau)$ on which the
limit is reached. Furthermore, dimensional analysis may be misleading
for frame components of the curvature. An example of this in spherical
symmetry is described below (a fan with $q<-1$), where a frame
component blows up at $\tau=-\infty$.

Because the natural dimension for an affine parameter is length,
dimensional analysis suggests that the affine parameter scales as
$e^{-\tau}$, so that DSS lines are finite as $\tau\to\infty$, but
infinitely extended as $\tau\to-\infty$. This is consistent with the
fact that $\tau=-\infty$ generically 
corresponds to infinite area radius $r$. We
therefore call $\tau=-\infty$ the ``kinematical infinity''. However,
Lake and Zannias \cite{LakeZannias} show that it is possible to have
complete null CSS lines. We have found an example of this in spherical
symmetry (a fan with $q=-1$), where the affine parameter is $\tau$,
rather than $e^{-\tau}$.

As we assume that the spacetime is {\it globally} fibrated by CSS or
DSS lines, its global manifold structure must be the product of a
3-manifold (with coordinates $x^i$) and the $\tau$ line, the cylinder
$M=\Sigma_3\times R$. Note that the geometry need not correspond to
the manifold structure. The kinematical singularity, for example, can be
a single point, a line, a 2-dimensional surface or a 3-dimensional
surface.

In addition to the kinematical singularity and infinity, singularities
and infinities can also be generated by the structure of $\bar
g_{\mu\nu}$. We shall call these ``dynamical''.  If we define a
singularity as an obstruction to continuing the spacetime, and
consider only inextendible spacetimes, by definition the boundaries of
the spacetime are either singularities or infinities. The kinematical
boundaries $\tau=\pm\infty$ are copies of $\Sigma_3$. The dynamical
boundaries of the spacetime must be invariant under the
self-similarity, and so the dynamical boundary has the manifold
structure $\partial \Sigma_3\times R$.

A hypersurface that is invariant under the self-similarity can in
general be deformed continuously into another such hypersurface. On
the other hand, this is generically not true for an invariant {\it
null} hypersurface. Such a hypersurface is therefore a geometrical
object and is called a self-similarity horizon (SSH). 


\section{Spherical symmetry}



\subsection{Spherical symmetry without self-similarity}


For the remainder of the paper we restrict our discussion to spherical
symmetry. The four-dimensional spacetime is the
product of a two-dimensional spacetime with coordinates $\tau$ and $x$
(the reduced spacetime) and a round two-sphere of area $4\pi r^2$. We
write the most general metric adapted to spherical symmetry in the
form
\begin{equation}
\label{spherical_metric}
ds^2=e^{-2\tau}\left(A\,d\tau^2+2B\,d\tau\,dx+C\,dx^2+F^2\,d\Omega^2
\right),
\end{equation}
where $d\Omega^2=d\theta^2+\sin^2\theta\,d\varphi^2$ is the metric on
the unit 2-sphere, and where $A$, $B$, $C$ and $F$ are functions of
$\tau$ and $x$. The explicit factor $e^{-2\tau}$ has been written in
anticipation of self-similarity. In the following we assume that the
signature is $(-,+,+,+)$, and that the metric is non-degenerate except
possibly at the boundaries, so that $AC-B^2<0$. Note that we do not
mean to imply that $\tau$ is timelike and $x$ spacelike: we have not
fixed the signs of $A$, $B$ and $C$. We choose $F\ge 0$ by
convention. For simplicity, we assume that $A,B,C,F$ are all $C^2$, so
that the Riemann tensor is $C^0$. (Note that a complete set of
Einstein equations can be constructed from only first derivatives of
$A,B,C$ and second derivatives of $F$.)

The area radius $r\equiv e^{-\tau}F$ is a scalar in the reduced
manifold. A second geometrical scalar is the Hawking mass $m$ defined
by $1-2m/r\equiv(\nabla r)^2$. From $m$ and $r$ we can define the
dimensionless scalar $\mu\equiv 2m/r$. A spherical surface (point in
the reduced spacetime) where $\mu\ge1$ is a closed trapped surface,
and one where $\mu=1$ is an apparent horizon.


\subsection{Spherical symmetry and self-similarity}


We now restrict to self-similarity, discussing the continuous and
discrete symmetries separately.

The metric (\ref{spherical_metric}) is CSS with homothetic vector
$\partial/\partial \tau$ if and only if $A$, $B$, $C$ and $F$ depend
only on $x$. (The same then holds for $\mu$.) The CSS lines
are the lines of constant $x$ on the reduced manifold. All coordinate
systems with this property are related by coordinate transformations
of the form
\begin{equation}
\label{cssgaugefreedom}
x'= \varphi(x),\qquad \tau'=\tau+\psi(x).
\end{equation}
The first of these just relabels the CSS lines. This is similar
to a change of radial coordinate in a coordinate system for
Schwarzschild (for example from the area radius to the isotropic
radial coordinate). The second changes the $\tau$ slicing, and can be
used to turn $\tau$ into a global coordinate. This is
similar to a change of slicing in Schwarzschild, for example from the
Schwarzschild to the Painlev\'e-G\"ullstrand time.

CSS lines are timelike for $A<0$, null for $A=0$, and spacelike for
$A>0$. Therefore the sign of $A$ is a geometric property of a
spacetime region. The lines of constant $\tau$ are timelike for $C<0$,
null for $C=0$, and spacelike for $C>0$. This is merely a property of
the coordinate system, and can be changed by a transformation of the
form (\ref{cssgaugefreedom}).

The metric (\ref{spherical_metric}) is DSS with scale period $\Delta$,
with the discrete conformal isometry
$\Phi:(\tau,x)\to(\tau+\Delta,x)$, if and only if
$A(\tau,x)=A(\tau+\Delta,x)$, and similarly for $B$, $C$ and $F$. (The
same then holds for $\mu$.) All coordinate systems adapted to
DSS in this way are related by coordinate transformations of the form
\begin{equation}
\label{dssgaugefreedom}
x'=\varphi(\tau,x),\qquad \tau'=\tau+\psi(\tau,x).
\end{equation}
where $\varphi$ and $\psi$ are periodic in $\tau$ with period $\Delta$. 

DSS lines can be moved around by a coordinate transformation (as
well as relabeled). It is clear that we cannot transform a DSS line
with $A>0$ for all $\tau$ 
into one with $A<0$, or vice versa. If the sign of $A$
changes with $\tau$ along a DSS line, the line can be moved
(``straightened out'') until $A$ is either strictly positive, strictly
negative or zero for all $\tau$. In the following we always assume
that the spacetime is covered by one global coordinate patch with this
property. This can always be done, although for numerical calculations
it may be better to use local patches in which the metric can
be simplified.

In such a coordinate system, the sign of $A$ is a geometric property
of a region of spacetime just as in CSS. This restriction does not
make the DSS lines with $A>0$ or $A<0$ rigid, because these are just
inequalities, but it does make the $A=0$ lines rigid: we cannot deform
such a line without making parts of it spacelike or timelike. $A=0$
lines are radial null geodesics that are invariant under the
self-similarity. In the full spacetime they correspond to spherical
self-similarity horizons. In the following, we shall focus on isolated
SSHs. When the SSH is not isolated, $A$ need not vanish there. An
example is given in Sec.~\ref{section:blocks}.

DSS lines and lines of constant $\tau$ are normal to each other if and
only if $B=0$. The geometric interpretation of the sign of $B$ is more
complicated, and will be deferred to Sec.~\ref{section:radialnull}.


\subsection{Boundaries of the reduced spacetime}


The 4-dimensional spacetime is fibrated by CSS lines or DSS lines, and
so the {\it reduced} spacetime is {\it foliated} by CSS or DSS lines,
all extending from $\tau=-\infty$ to $\tau=\infty$. The manifold
structure of the reduced spacetime is therefore a rectangle, the
$\tau$ line times an interval in $x$. In the DSS case, we make the
assumption that both dynamical boundaries, as well as any SSHs, are
DSS lines $x=\rm const$. These are only local restrictions on the
coordinate system. We then have two
kinematical boundaries $\tau=\pm\infty$ and two
dynamic boundaries $x=x_{\rm min,max}$. There are three types of
dynamical boundary:

A dynamical infinity is a dynamical boundary where radial geodesics
reach infinite affine parameter with finite $\tau$.
Most naturally this coincides with $F=\infty$, but finite
$F$ is also conceivable, corresponding to a wormhole with topology
$R^2\times S^2$ that stretches to infinity in the $x$-direction with
finite circumferences $2\pi r$.

A dynamical singularity is a dynamical boundary where radial null
geodesics end in infinite curvature at finite affine parameter, at
finite $\tau$. A dynamical singularity can be central, with $F=0$, or
conceivably be non-central, with $F\ne 0$. As mentioned in the
discussion of the general (non-spherical case), any weak singularities
through which the spacetime can be continued would be considered
interior points.

A regular center is given by $F=0$ with the additional condition $\mu\sim
F^2$ as $F\to 0$. This is equivalent to $m\sim r^3$, or the absence
of a defect angle. This boundary of the reduced spacetime is of course
not a boundary of the full spacetime but just the central world line.

$F(x,\tau)=0$ or $\infty$ always corresponds to one of these three
boundaries, and so cannot occur except at the dynamical
boundaries. Therefore, as $r=e^{-\tau}F(\tau,x)$, the kinematical
boundary $\tau=\infty$ is always central and the kinematical boundary
$\tau=-\infty$ always has infinite area. However, infinite area radius
$r$ does not necessarily mean an infinity in the sense of infinite
affine parameter, as we shall see below in
Sec.~\ref{section:singularfan}.


\section{Self-similarity horizons}
\label{section:SSH}



\subsection{Radial null geodesics}
\label{section:radialnull}


In this section we investigate the spacetime structure near isolated
self-similarity horizons in spherical symmetry. Without loss of
generality, we assume in the following that the isolated SSH under
consideration is at $x=0$.

The radial null geodesics of the metric (\ref{spherical_metric}) are
given by
\begin{equation}
\label{f+-}
{dx\over d\tau} = C^{-1}\left(-B \pm\sqrt{B^2-AC}\right)\equiv f_\pm(x,\tau).
\end{equation}
If we cover the entire spacetime with a global coordinate patch, then
the two signs in the equation above correspond to the two global
families of null geodesics that cross-hatch the reduced spacetime. We
shall refer to these as $+$ and $-$ lines. In the conformal diagram
they are the left and right moving null lines (where of course left
and right are purely conventional names).

The radial null geodesics can also be written as
\begin{equation}
\label{fuv}
{dx\over d\tau} = C^{-1}\left(- B  \pm {B\over |B|}\sqrt{B^2-AC}\right)\equiv
f_{u,v}(x,\tau).
\end{equation}
We refer to first of these as lines of constant $u$, and the second as
lines of constant $v$. In the limit $A\to 0$ these two families become
\begin{eqnarray}
\label{ulines}
{dx\over d\tau} &=& f_u =  -{A\over 2B}\left[1+O(AC/B^2)\right], \\
\label{vlines}
{d\tau\over dx} &=& {1\over f_v} = -{C\over 2B}\left[1+O(AC/B^2)\right].
\end{eqnarray}
Note that as $A\to 0$, we must have $B\ne 0$, because in any regular
global coordinate system we have $B^2-AC>0$. The SSH $x=0$ is clearly
a $u$ line (solution of the upper equation), and so the $u$ lines are
parallel to this particular horizon. Without loss of generality,
we set $u=0$ on the SSH. The $v$ lines cross the horizon.

The classification of radial null geodesics as $+$ and $-$ lines is
global, and the classification as $u$ and $v$ lines is local to a
SSH. Their relation depends on the sign of $B$: for $B>0$, the $u$
lines are the $+$ lines, while for $B<0$, they are the $-$
lines. Therefore if $B$ does not change sign between two isolated
SSHs, they are parallel in the sense that both are $+$ or both are
$-$. In the other case, a neigbouring $+$ and $-$ SSH cannot intersect
at finite values of their affine parameters, but they can meet at
their endpoints, thus forming a ``corner'' in the reduced
spacetime. This idea will be taken up again in
Sec.~\ref{section:blocks}.


\subsection{SSHs as attractors}


The SSH $x=0$ is a stationary point of the dynamical system
$dx/d\tau=f_u$. For $xf_u(x)<0$, $|x|\to 0$ as $\tau\to\infty$, while
for $xf_u(x)>0$, $|x|\to 0$ as $\tau\to-\infty$.  If the metric is
$C^1$ at the SSH, then $A=O(x)$, and therefore $f_u=O(x)$. To see that
the attracting fixed point $x=0$ is actually reached, note that
(assuming $C^2$ for simplicity)
\begin{equation}
f_u(\tau,x)\equiv f_1(\tau) x+O(x^2), \qquad f_1(\tau)\equiv\bar f_1+\tilde
f_1(\tau), 
\end{equation}
where $\bar f_1$ is the average value of the periodic function
$f_1(\tau)$. Then 
\begin{equation}
\left[1+O(x)\right]\ln x= \bar f_1\tau+\int \tilde f_1(\tau)\,d\tau,
\end{equation}
where the integral is periodic. A similar calculation holds if the
leading order in $f$ is higher than $O(x)$.

We call the SSH a ``splash'' if it attracts $u$ lines as
$\tau\to\infty$, and a ``fan'' if it attracts them as
$\tau\to -\infty$. The motivation for these names will become clear
below when we investigate the spacetime structure near the SSH. If
$f_u(x)$ has the same sign on both sides of $x=0$, the SSH attracts
$u$ lines as $\tau\to\infty$ from one side, and as $\tau\to-\infty$
from the other. We then call it half a fan from one side and half a
splash from the other.

In a splash, the coordinate location $(x=0,\tau=\infty)$ is
intersected by a range of $u$ lines, and so it cannot be a point, but
must be a line in the reduced spacetime. The same is true for the
coordinate location $(x=0,\tau=-\infty)$ in a fan.


\subsection{Double null coordinates}


We now show that the line $x=0,\tau=\pm\infty$ is null by explicitly
constructing double null coordinates on the reduced spacetime.  We
assume that the metric is regular and (for simplicity) $C^2$ at the
SSH and that the SSH is isolated. The metric near the SSH then has the
form
\begin{eqnarray}
\nonumber
ds^2 &=&e^{-2\tau} \Bigl[\left(a(\tau)x+O(x^2)\right)\,d\tau^2
+2\left(b(\tau)+O(x)\right)\,d\tau\,dx\\
\label{approx}
&&+\left(c(\tau)+O(x)\right)\,dx^2
+\left(f(\tau)^2+O(x)\right)\,d\Omega^2\Bigr].
\end{eqnarray}
In Appendix~\ref{appendix:SSH} we show that by a coordinate
transformation of the type (\ref{dssgaugefreedom}) we can bring the
metric into the unique form where $a(\tau)=4q$, $b(\tau)=1$ and
$c(\tau)=0$. Then
\begin{eqnarray}
\nonumber
ds^2 &=&e^{-2\tau} \Bigl[\left(4qx+O(x^2)\right)\,d\tau^2
+\left(2+O(x)\right)\,d\tau\,dx\\
\label{standardapprox}
&&+O(x)\,dx^2
+\left(f(\tau)^2+O(x)\right)\,d\Omega^2\Bigr].
\end{eqnarray}
Note that $f_u\simeq -2qx$ as $x\to 0$ and so the SSH is a splash for
$q>0$ and a fan for $q<0$.

We now define a new spacetime, the ``q-metric'', by dropping all
higher order terms from the approximation:
\begin{equation}
\label{q-metric}
ds^2 = e^{-2\tau} \left(4qx\,d\tau^2+2\,d\tau\,dx+f(\tau)^2\,d\Omega^2\right).
\end{equation}
Its region $|x|<\epsilon$ for some small constant $\epsilon$ is an
approximation to a generic spherically symmetric DSS spacetime near an
isolated SSH, but the q-metric is simple enough to be studied exactly.

For $q\ne 0,-1$, we define a pair of null coordinates $u$ and $v$ by
\begin{eqnarray}
\label{genericuv}
u&=&{1\over p}xe^{2q\tau}, \\
v&=&e^{-2p\tau},
\end{eqnarray}
where $p\equiv q+1$. The metric in coordinates $u$ and $v$ is
\begin{equation}
\label{nullform}
ds^2=-du\,dv 
+f^2(\tau)\,v^{1/p} \,d\Omega^2.
\end{equation}
The definitions of $u$ and $v$ can be inverted to give
\begin{eqnarray}
\label{xuv}
x&=& pu v^{q/p}, \\
\tau&=&-{\ln v\over 2p}.
\end{eqnarray}
The range of coordinates in the q-metric is $-\infty<u<\infty$ and
$0<v<\infty$. As an approximation to generic SSHs, this metric is good
only for $|x|<\epsilon$, which translates to
\begin{equation}
|u|<{\epsilon\over p}v^{-q/p}.
\end{equation}

In order to draw the conformal diagram, we compactify the metric
(\ref{nullform}) by setting $v=(\tan V)^{-p/q}$ and $u=(x_0/p)\tan U$
for a constant $x_0$. This gives 
\begin{eqnarray}
ds^2 &=& \omega^2 \left({x_0\over q} dU\,dV +
f(\tau)^2 \sin^2 V\cos^2 U\,d\Omega^2\right), \\
\omega^2&=&(\sin V)^{-2-1/q}(\cos V)^{1/q}(\cos U)^{-2}
\end{eqnarray}
for $0\le V\le \pi/2$ and $-\pi/2\le U\le \pi/2$. With
\begin{equation}
\label{xuV}
x=x_0{\tan U\over \tan V},\qquad \tau={1\over 2q}\ln\tan V
\end{equation}
$x=0$ has two branches: $U=0$ and $V=\pi/2$. Therefore the SSH $x=0$
is bifurcate in the form of a T. $U=\pm V$ corresponds to $x=\pm x_0$,
for example the limits $x=\pm \epsilon$ of the validity of our
approximation. 


\subsection{Geodesics near the SSH}


To understand the geometry of the $V=\pi/2$ branch of the SSH, we now
investigate the behaviour of the area radius $r$, the affine parameter
and the spacetime curvature along geodesics running into $V=\pi/2$.

Like the spacetimes it approximates, the q-metric is exactly
self-similar. In the CSS case the homothetic vector is
\begin{equation}
{\partial\over\partial \tau}=2qu {\partial\over\partial u}
-2pv{\partial\over\partial v}.
\end{equation}
However, the q-metric also has the exact Killing vector
${\partial/\partial u}$, which a generic SSH does not have. The
presence of this Killing vector means also that the SSH of the
q-metric is not isolated, as every line $u=\rm const$ is a SSH.

Without loss of generality we assume that the geodesics are in the
equatorial plane. Spherical symmetry and the null Killing vector give
rise to the conserved quantities
\begin{equation}
L\equiv \left({\partial\over\partial \varphi}\right)^a u_a =f^2
v^{1/p}\ \dot \varphi, \qquad E\equiv -2\left({\partial\over\partial
u}\right)^a u_a=\dot v,
\end{equation}
where $u^a$ is the tangent vector to the geodesic, and a dot denotes
the derivative with respect to the affine parameter. The norm
\begin{equation}
\kappa\equiv u^au_a=-\dot u\dot v+f^2(\tau)v^{1/p}\dot\varphi^2=0,1,-1
\end{equation}
is also conserved. As $\dot v=E$ is constant, $v$ is an affine
parameter for all 4-dimensional geodesics, except for the $v=\rm
const$ radial null geodesics. Defining the impact
parameter $D\equiv L/E$, we can integrate the geodesics to give
\begin{eqnarray}
\label{nonradial}
\varphi-\varphi_0&=&D\int_0^vf^{-2}v^{-1/p}\,dv
\simeq {pD\over qf^2}\,v^{q/p}, \\
\nonumber
u-u_0&=&D^2\int_0^vf^{-2}v^{-1/p}\,dv-{\kappa\over  E^2}\,v \\
&\simeq&{pD^2\over qf^2}\,v^{q/p}-{\kappa\over  E^2}\,v,
\end{eqnarray}
where the second equality in each case holds exactly in the CSS case,
where $f(\tau)$ is constant, and approximately in the DSS case.

In the CSS case, the homothety gives rise to the quantity
\begin{equation}
\label{Jdef}
J\equiv \left({\partial\over\partial \tau}\right)^a u_a=-qu\dot v
+pv\dot u.
\end{equation}
It is straightforward to check that for all geodesics this correctly
obeys
\begin{equation}
\dot J=-\kappa.
\end{equation}
Therefore it was consistent to use the Killing symmetry, rather than
the homothetic symmetry, in order to integrate the geodesic
equations. 


\subsection{Curvature} 


From the results of Appendix~\ref{appendix:curvature}, the components
of the 4-dimensional Ricci tensor in a parallely propagated null
tetrad and the Ricci scalar of the q-metric (for $q\ne 0,-1$) are
\begin{eqnarray}
\label{R22}
R_{22}&=&{2q+1\over 2p^2}v^{-2}, \\
R_{33}&=&R_{44}=r^{-2}, \\
R&=&2r^{-2}.
\end{eqnarray}
Therefore, $r=0$ is a curvature singularity for all values of $q$, and
$v=0$ is a curvature singularity for $q\ne-1/2$. Because
\begin{equation} 
v=(\tan V)^{-{p\over q}}, \qquad r=f(\tau)\,(\tan V)^{-{1\over 2q}}, 
\end{equation}
for $q>0$ there is a curvature singularity at $V=\pi/2$, for $-1<q<0$
there is a curvature singularity at $V=0$, and for $q<-1$, there are
curvature singularities at both $V=0$ and $V=\pi/2$. In this last
case, the curvature singularity at $V=\pi/2$ is non-scalar at
$r=\infty$ and is not predicted by dimensional analysis: $R$ vanishes
there, but $R_{22}$ blows up.

The 4-dimensional Riemann tensor of any spherically symmetric
spacetime is constructed from the 2-dimensional Riemann tensor of the
reduced spacetime, plus derivatives of $r$. Any curvature singularity
in the q-metric that arises from derivatives of $r$ has a counterpart
in the SSH it approximates. However, we must keep in mind that the
2-dimensional curvature of the reduced spacetime is generically
nonzero, and blows up at $\tau=\infty$, while the q-metric
approximates it as zero. But we have already identified $\tau=\infty$ as
a curvature singularity because $R\sim r^{-2}$.


\subsection{Nature of the boundary $V=\pi/2$} 


We now have all the tools we need to characterise the neighbourhood
$|x|<\epsilon$ of the SSH when $q,p,f(\tau)\ne 0$. The special cases
$q=-1$, $q=0$ and $f(\tau)=0$ need to be considered separately.


\subsubsection{Generic case $q>0$: splash} 

\begin{figure}
\includegraphics[width=\columnwidth]{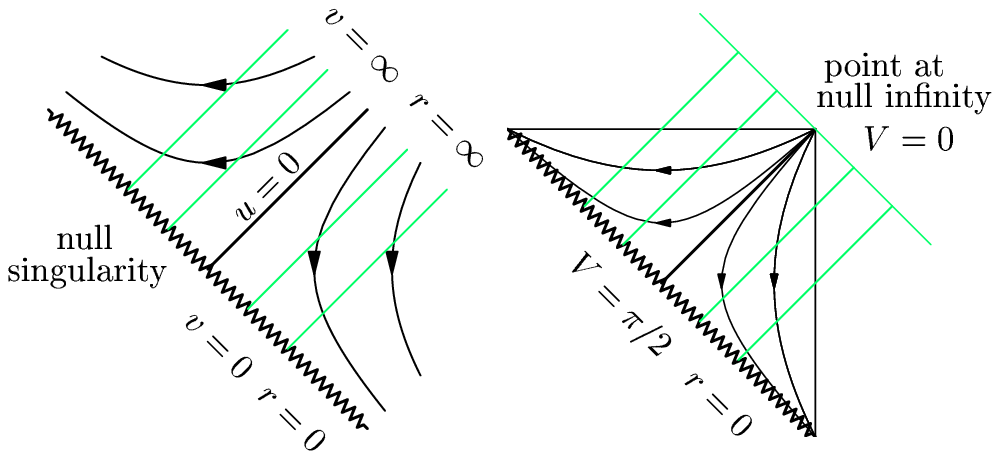}
\caption{
A splash. The conformal diagram on the right has been
obtained from the one on the left by compactification. The lines with
arrows are DSS lines with the arrow pointing towards increasing
$\tau$. The thin parallel lines are radial null geodesics $u={\rm
const}$. The central one $u=0$ is also the DSS line $x=0$.  
}
\label{figure:splash}
\end{figure}

$V=\pi/2$ corresponds to $(x=0,\tau=\infty)$, $r=0$ and $v=0$. It is a
null central scalar curvature singularity, reached by all geodesics
except the radial null geodesics $v=\rm const$ at finite affine
parameter. The values of $u$, $\theta$ and $\varphi$ also converge as
the singularity is reached, and so the singularity is a 3-dimensional
null surface, rather than a line, in the 4-dimensional spacetime. A
splash is illustrated in Fig.~\ref{figure:splash}.

\subsubsection{Generic case $-1<q<0$: regular fan} 

\begin{figure}
\includegraphics[width=\columnwidth]{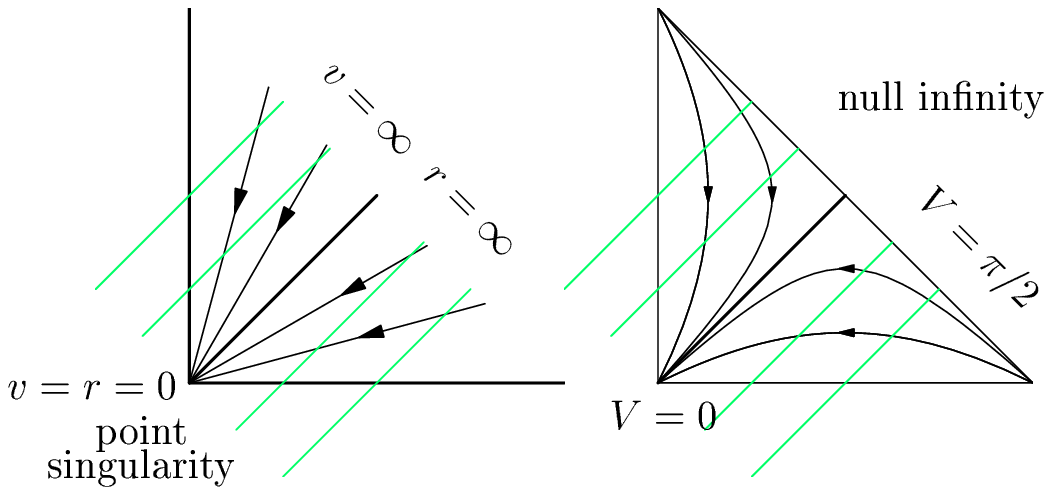}
\caption{ A fan. The conformal diagram on the right has been obtained
from the one on the left by compactification. The lines with arrows
are DSS lines with the arrow pointing towards increasing
$\tau$. The thin parallel lines are radial null geodesics $u={\rm
const}$. The central one $u=0$ is also the DSS line $x=0$.  }
\label{figure:fan}
\end{figure}

$V=\pi/2$ corresponds to $(x=0,\tau=-\infty)$, $r=\infty$ and
$v=\infty$.  It is reached by all null geodesics at infinite affine
parameter $v$ and finite values of $u$, $\theta$ and $\varphi$. It is
missed by all spacelike and timelike geodesics. All curvature scalars
and frame components vanish there. In this sense it has the same
structure as the null infinity of Minkowski spacetime. A regular fan
is illustrated in Fig.~\ref{figure:fan}.

\subsubsection{Generic case $q<-1$: singular fan} 
\label{section:singularfan}


$V=\pi/2$ corresponds to $(x=0,\tau=-\infty)$, $r=\infty$ and
$v=0$. The curvature scalars vanish but the frame component $R_{22}$
blows up there, and all geodesics end at finite affine parameter. This
part of the ``kinematical infinity'' is therefore in fact a
(non-scalar) curvature singularity. From (\ref{R22}) we see that
$R_{22}$ is negative for $q<-1/2$. As $R_{11}$ vanishes and the other
frame components can be neglected as $x\to 0$, the energy density is
therefore negative for any timelike observer. Therefore, although it
is kinematically allowed, this spacetime cannot arise with reasonable
matter content.


\subsubsection{Special case $q=-1$: regular fan}


This implies $p=0$ and so must be treated specially. Double null
coordinates are now
\begin{eqnarray}
u&=&xe^{-2\tau}, \\
v&=&-2\tau
\end{eqnarray}
and the q-metric is
\begin{equation}
ds^2= -du\,dv+e^vf^2d\Omega^2.
\end{equation}
The inverse coordinate transformation gives
\begin{equation}
x=ue^{-v}, \qquad
\tau = -{v\over 2},
\end{equation}
and so $x\to 0$ along radial null geodesics $u={\rm const}$ as
$\tau\to -\infty$, which is also $r\to \infty$ and $v\to \infty$. The
analysis of non-radial, non-null geodesics gives (assuming CSS for
simplicity) gives
\begin{equation}
\label{nonradial2}
\varphi-\varphi_0=-{D\over f^2}\,e^{-v}, 
\qquad
u-u_0=-{D^2\over f^2}\,e^{-v}-{\kappa\over  E^2}\,v.
\end{equation}
Therefore $v=\infty$ has the same structure as Minkowski null
infinity, and so the SSH is a regular fan. The only qualitative
difference to the $-1<q<0$ fan is that the outgoing radial null
geodesic $u=0$ is infinitely extended at both ends.


\subsubsection{Special case $q=0$: regular fan, splash or half of each}

 
We have assumed that $A(\tau,x)$ is generic in the sense that near
$x=0$, $A=a(\tau)x+O(x^2)$ with $a\ne 0$. The special case
$a(\tau)=0$ and hence $q=0$ corresponds to the $O(x)$ term being
absent. As we are investigating isolated SSHs, $A=0$ is not a
sufficiently good approximation, and we must include the first
non-vanishing order in $A$. As an example, we consider
\begin{equation}
f_u(x,\tau)\equiv f_m(\tau) x^m+O(x^{m+1}), \qquad f_m(\tau)\equiv\bar
f_m+\tilde f_m(\tau),
\end{equation}
where $m=1$ is the generic case that we considered before, and we now
consider $m>1$. Then 
\begin{equation}
-\left[1+O(x)\right]{x^{-(m-1)}\over m-1}\simeq \bar f_m\tau+\int
 \tilde f_m(\tau)\,d\tau.
\end{equation}
Consider the case where $m>1$ is an integer. If it is odd, $x\to 0$ as
$f_m\tau\to \infty$, and so we have a fan for $f_m<0$ and a splash for
$f_m>0$. If $m$ is even, then $x\to 0$ as ${\rm sign}(x)f_m\tau\to
\infty$, and so we have half a fan on one side and half a splash on
the other. 

To see if the affine parameter on $u$ lines is finite or infinite as
$x\to 0$, we assume CSS for simplicity. From the definition
(\ref{Jdef}) of $J$, the form (\ref{spherical_metric}) of the metric,
and the approximation (\ref{ulines}) to the $u$ lines, we find
\begin{equation}
J\simeq - e^{-2\tau}B(x)\dot x.
\end{equation}
On null geodesics $J$ is conserved. As the metric is regular at $u=0$,
$B$ cannot vanish, and so is approximately constant as $x\to 0$. We
then have
\begin{equation}
\label{f=0}
{d\lambda\over d\tau} \sim e^{-2\tau}{dx\over d\tau} \sim e^{-2\tau} 
|\tau|^{-{m\over m-1}}
\end{equation}
For all $m>1$, $\lambda(\tau)$ converges as $\tau\to\infty$ and
diverges as $\tau\to-\infty$. Therefore the affine parameter $\lambda$
converges at $x=0$ if and only if $r\sim e^{-\tau}$ converges. This
means that the SSH is either a splash or a regular fan, but never a
singular fan. Note that for a dynamical boundary which is a null
infinity, the metric cannot have the form (\ref{f=0}). An example is
found in App.~\ref{appendix:brady}.


\subsubsection{Special case $f(\tau)=0$}


Until now we have assumed that $F(x,\tau)$ does not vanish or blow up
at the horizon. Even if it does, our analysis of double null
coordinates and radial geodesics in the q-metric is unchanged. For
simplicity we assume CSS again, so that $F=F(x)$, and as an example we
consider $F(x)\simeq f_n x^n$, where $n$ is not necessarily an
integer. The q-metric becomes
\begin{equation}
ds^2= -du\,dv + f_n^2\,(puv^{q/p})^{2n}\,v^{1/p} \,d\Omega^2.
\end{equation}
The branch $u=0$ of the SSH now has $r=0$ for $n>0$ and $r=\infty$ for
$n<0$.  The curvature of this metric is given in appendix $B$. The
result is that the branch $u=0$ of the SSH is a curvature singularity
for all $n\ne 0$. In our terminology, the SSH is a dynamical
singularity and forms one end of the range of $x$.


\section{Fans and splashes as building blocks}
\label{section:blocks}


We now show how the conformal diagram of the reduced spacetime can
be constructed from fans and splashes. We shall refer to $+$ and $-$
fans and splashes, according to if the $u$ lines are the $+$ or the
$-$ lines. 

\begin{figure}[t]
\includegraphics[width=7cm]{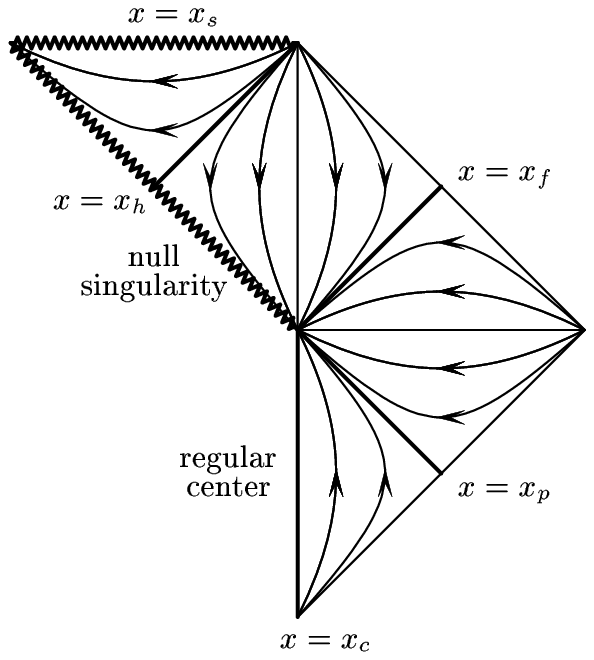}
\caption{Penrose diagram for a self-similar spacetime with a regular
center at $x=x_c$, two fans at $x=x_p$ and $x=x_f$, and one splash at
$x=x_h$. Because the spacetime contains at least one splash, the
kinematical singularity is extended and null. Because it contains just
one splash, the dynamical singularity $x=x_s$ is spacelike.  This
spacetime occurs in CSS with scalar field matter, compare Fig. 22 of
\cite{OriPiran2}.}
\label{figure:spacelikesingular}
\end{figure}

Consider two neighbouring isolated SSHs $x_1$ and $x_2$, so that $A$
vanishes for $x=x_1$ and $x=x_2$ and is either strictly positive or
strictly negative in between. If $B$ does not change sign between the
two SSHs, then they are one fan and one splash, either both $+$ or
both $-$. An example of this appears near the top of
Fig.~\ref{figure:spacelikesingular}. There $x_f$ is a $+$ fan and
$x_h$, to the future of $x_f$, is a $+$ splash. If $B$ does change
sign between two neighbouring SSHs, then they are either two fans or
two splashes. One is $+$ and one is $-$.  An example of this appears
toward the right of Fig.~\ref{figure:spacelikesingular}. $x_p$ is a
$-$ fan, to the past of the $+$ fan $x_f$.

As every SSH is an attractor of one type of radial null geodesic in
one direction, and as there are no other fixed points of $dx/d\tau=f$,
almost all points on the kinematical boundaries are part of a
fan or splash, and so all have the same value of $x$. In
Fig.~\ref{figure:spacelikesingular} these values are $x_f$, $x_p$ and
$x_h$. By definition, the two dynamical boundaries are also at
constant values of $x$. In Fig.~\ref{figure:spacelikesingular} these
are $x_c$ and $x_s$, a regular center and a singularity. All
intermediate values of $x$ are bunched up in isolated points on the
boundary of the conformal diagram.

Because the kinematic boundaries of the conformal diagram are
provided by the fans and splashes, all kinematically possible Penrose
diagrams of self-similar spacetime diagrams can be enumerated by
stringing together, like dominoes, fans and splashes (or sometimes
half-fans and half-splashes). As basic building blocks one could pick
the half-fan and the half-splash, and one could formalize a set of
rules that govern how these can be combined.

\begin{figure}[t]
\includegraphics[width=6cm]{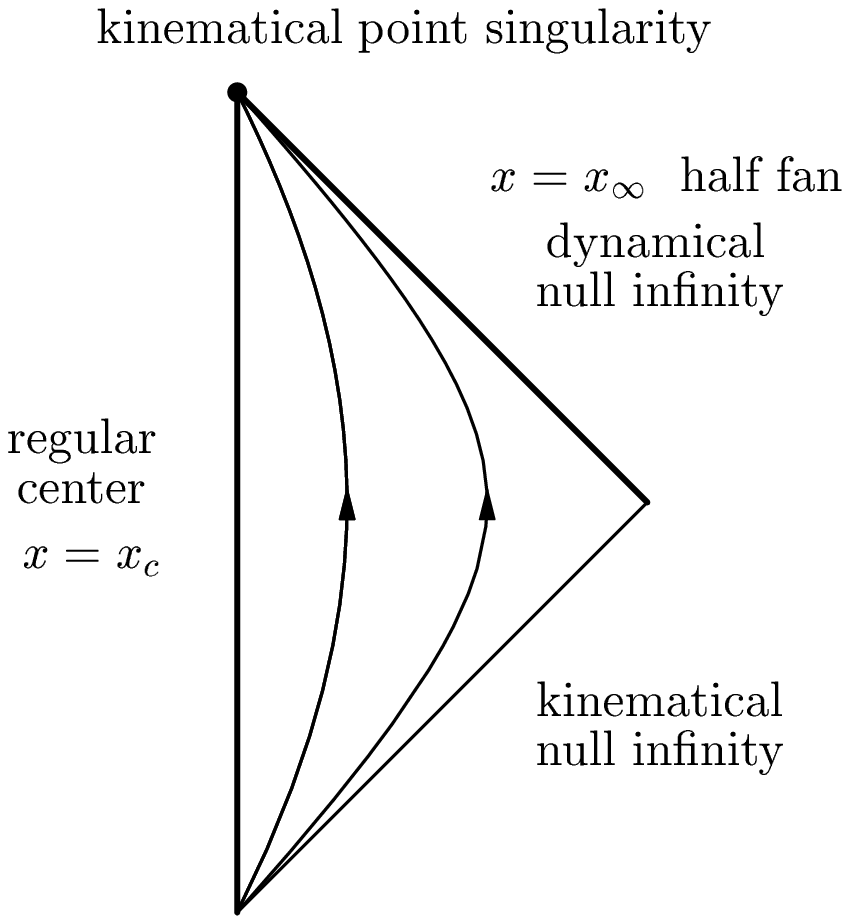}
\caption{Penrose diagram for a self-similar spacetime with a regular
center at $x=x_c$, and half a fan which is also a dynamical (null)
singularity at $x=x_\infty$. This spacetime occurs in CSS with scalar
field matter, Class I of \cite{Brady}, with coordinates $x_c=0$ and
$x_\infty=\infty$. The metric is given in App.~\ref{appendix:brady}.}
\label{figure:brady}
\end{figure}

\begin{figure}[t]
\includegraphics[width=7cm]{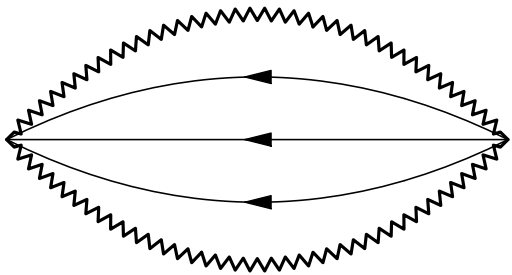}
\caption{Penrose diagram for a spherically symmetric solution without
any SSH. Here the kinematical infinity and kinematical singularity each consist
of one point in the reduced spacetime. The rest of the boundary
consists of two spacelike (central) dynamical singularities (both
linked to the kinematical singularity). This spacetime can be realized
dynamically with perfect fluid matter \cite{fluidcss}.}
\label{figure:clam}
\end{figure}

While the example in Fig.~\ref{figure:spacelikesingular} gives clear
examples of fans and splashes, they are harder to spot in spacetimes
with simpler conformal diagrams. We give two examples of this.
Fig.~\ref{figure:brady} contains only a single building block, a
half-fan. Its SSH is at the same time a dynamical infinity.
Fig.~\ref{figure:clam} illustrates a spacetime with no fans or
splashes. The kinematical infinity and singularity are each reduced to
a point, and the dynamical boundaries are both singularities.

\begin{figure}[t]
\includegraphics[width=7cm]{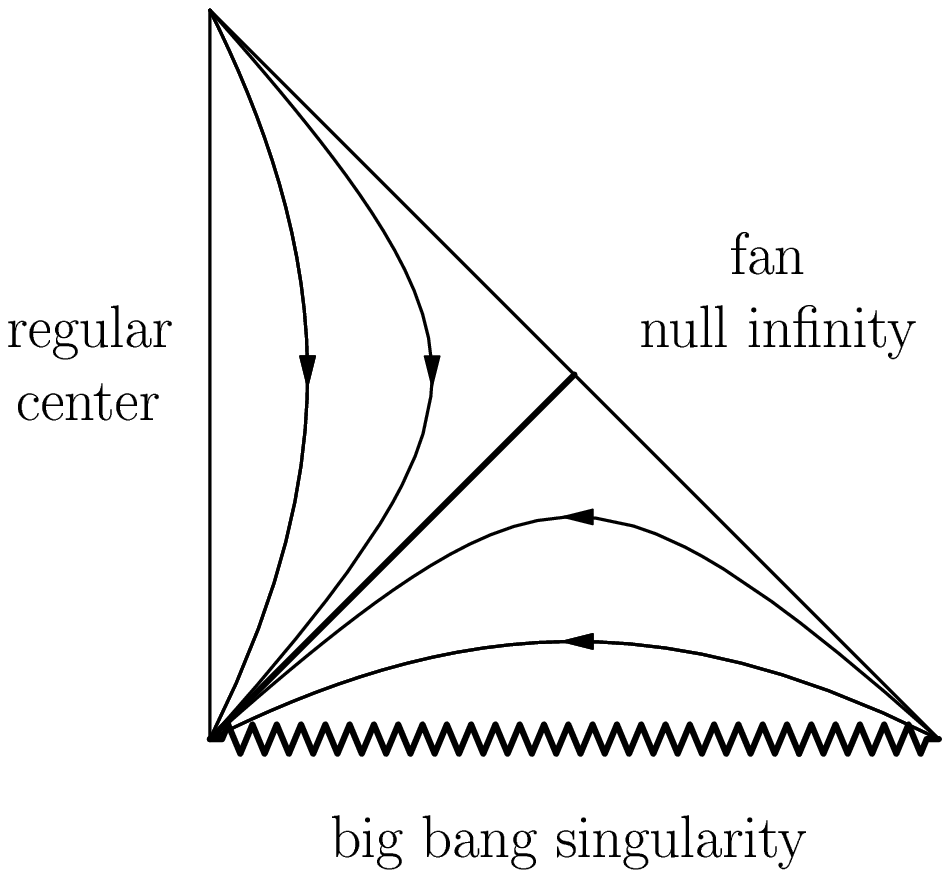}
\caption{Penrose diagram for the flat Friedmann solution. Note that
because of translation invariance, any point in space can be chosen as
the regular center.}
\label{figure:flatfriedmann}
\end{figure}

If a spherical CSS spacetime has additional symmetries,
our classification in terms of fans and splashes is still applicable,
but no longer unique (or very useful). We give two examples.
Fig.~\ref{figure:flatfriedmann} illustrates the flat Friedmann
solution. It is CSS for any perfect fluid matter with the linear
equation of state $p=c_s^2\rho$. This is shown by writing the metric
in the form
\begin{equation}
ds^2=e^{-2\tau} \left[(-1+\eta^2 x^2)d\tau^2
-2\eta x\,dx\,d\tau+dx^2+x^2\,d\Omega^2\right], 
\end{equation}
where $\eta=(1+3c_s^2)/3(1+c_s^2)$. Clearly there is a fan at
$x=1/\eta$. The big bang is a dynamical singularity in our
classification, with the exception of the point at the origin of
spherical symmetry, which is the kinematical singularity. But because
the solution has additional $E(3)$ translation invariance, every fluid
world line can be considered as the center of spherical symmetry. The
distinction between the kinematical and dynamical singularity is then
completely artificial. For the same reason there is a spherical SSH
through every point of the spacetime.

\begin{figure}[t]
\includegraphics[width=7cm]{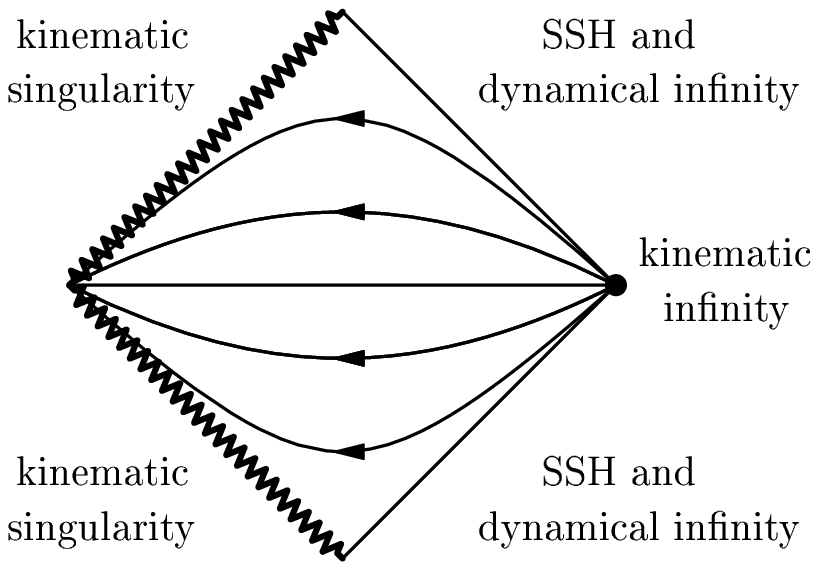}
\caption{Penrose diagram for the Hayward solution. The integral curves
of the homothetic vector $\partial/\partial \tau$ are shown, but these
CSS lines are not unique because the spacetime has two linearly
independent homothetic vector fields.}
\label{figure:hayward}
\end{figure}

Fig.~\ref{figure:hayward} illustrates the scalar field solution
discussed by Hayward \cite{Hayward}, with the metric
\begin{equation}
ds^2 = e^{-2\tau} [ 2\,d\tau^2 -2\,dx^2 + d\Omega^2 ]
\end{equation}
and scalar field $\sqrt{4\pi G}\phi=x$. The spacelike homothetic
vector $\partial/\partial \tau$ is evident. However,
$\partial/\partial x$ is a timelike Killing vector, and any linear
combination of $\partial/\partial \tau$ and
$\partial/\partial x$ with constant coefficients 
is also a homothetic vector. In particular, the
two linear combinations $\partial/\partial x\pm \partial/\partial
\tau$ are null homothetic vectors. Therefore, there are no isolated
SSHs, and two SSHs cross in every point of the reduced spacetime. In
Fig.~\ref{figure:hayward} we have marked the CSS lines of
$\partial/\partial \tau$. In that view, the spacetime diamond is made up from
two disjoint half-splashes.

Recall that $A=0$ at any isolated SSH in CSS. The Hayward spacetime
shows that this need not be true if the SSH is not isolated. The
reason is that now CSS lines with arbitrary directions go through
every point. None of these directions is preferred, but of course only
one is null.


\section{Conclusions}


After a brief discussion of geometric self-similarity in general, we
restricted attention to spacetimes that are both spherically symmetric
and self-similar. Our investigation has been kinematical in that it uses
only those symmetries but not the Einstein equations.

As is well-known, every spherical spacetime can be written as the
product of a reduced 1+1-dimensional spacetime with a spacelike
2-sphere at each point. Continuous self-similarity provides a
preferred foliation of the reduced spacetime by the integral curves of
the homothetic vector field, or CSS lines. In DSS, a foliation by DSS
lines is not unique, but it can be restricted enough to serve
essentially the same purpose. This, combined with dimensional
analysis, suggested a heuristic classification of the boundaries of
the reduced spacetime into a kinematical infinity $\tau=-\infty$ where
all DSS lines begin, a kinematical singularity $\tau=\infty$ where
they end, and two dynamical boundaries $x=x_{\rm min}$ and $x=x_{\rm
max}$ which {\it are} DSS lines. Each of the two dynamical boundaries
can be either a regular center, a singularity, or an infinity.

If this picture is correct, all singularities are connected: the
kinematical singularity, plus possibly a dynamical singularity
connected to it at either end. Similarly, all infinities are
connected: the kinematical infinity, plus possibly a dynamical
infinity connected to it at either end. This would rule out spacetimes
with two or more disjoint singularities, for example cosmologies with
a big bang and big crunch, or two or more infinities, for example
Kruskal-like spacetimes \cite{fluidcss}.

We then focussed on SSHs: radial null geodesics that are invariant
under the self-similarity. We investigated and classified a fairly
generic family of SSHs by approximating them by a class of spacetimes
we call the q-metric. We showed that these SSHs are not simply lines
in the interior of the conformal diagram, but are T-shaped, with the
crossbar of the T forming part of the either the kinematical infinity
(fan) or kinematical singularity (splash). A neighbourhood of the SSH
is therefore a triangle in the conformal diagram. The entire conformal
diagram of a generic spherical DSS spacetime can be assembled from
these extended SSHs. We have attempted a general classification, but
our analysis is not rigorous, and we may have overlooked some
kinematical possibilities. At least our framework fits all spherical
DSS solutions known to us including some degenerate cases.

A problem arises in the q-metric with $q<-1$, where a part of what by
dimensional analysis should be the kinematical infinity is in fact a
non-scalar curvature singularity. However, these spacetimes have
negative energy density, and imposing an energy condition saves the
``one singularity, one infinity'' result suggested by dimensional
analysis.

 
\acknowledgments We would like to thank James Vickers and Bernard
Carr for helpful discussions. This research was supported by EPSRC
grant GR/N10172/01.
 

 
\appendix


\section{Coordinate transformation near the SSH}
\label{appendix:SSH}


Here we show how the standard form (\ref{standardapprox}) of the
metric near a SSH can be obtained. We assume that we have already used
part of the gauge freedom (\ref{dssgaugefreedom}) to set $x=0$ along
the SSH in order to obtain (\ref{approx}). We now carry out a further
coordinate transformation of the form (\ref{dssgaugefreedom}) which
preserves this form of the metric, with new coefficients $a'(\tau)$,
$b'(\tau')$, $c'(\tau')$ and $f'(\tau')$. We expand
(\ref{dssgaugefreedom}) and impose the condition that $O(x)=O(x')$. To
leading order this gives
\begin{equation}
x'=\varphi_1(\tau) x+O(x^2), 
\quad \tau'=\tau+\psi_0(\tau) +\psi_1(\tau) x+O(x^2)
\end{equation}
Comparing coefficients, we find that
\begin{eqnarray}
a&=&e^{-2\psi_0}\left[
\varphi_1(1+\dot\psi_0)^2 a'+2\dot \varphi_1(1+\dot\psi_0) b'\right], \\
b&=&e^{-2\psi_0}\varphi_1(1+\dot\psi_0)b', \\
c&=&e^{-2\psi_0}\left[\varphi_1^2 c'+ 2\varphi_1\psi_1 b'\right], \\
f&=&e^{-\psi_0}f'
\end{eqnarray}
where $\dot \varphi_1=d\varphi_1/d\tau$ and
$\dot\psi_0=d\psi_0/d\tau$. We now impose the gauge conditions
\begin{equation}
a'=4q,\qquad b'=1, \qquad c'=0,
\end{equation}
where $q$ is a constant to be determined. This gives the coupled ODEs
\begin{eqnarray}
\label{1}
a&=&e^{-2\psi_0}\left[
\varphi_1(1+\dot\psi_0)^2 4q+2\dot \varphi_1(1+\dot\psi_0) \right], \\
\label{2}
b&=&e^{-2\psi_0}\varphi_1(1+\dot\psi_0) 
\end{eqnarray}
for $\psi_0$ and $\varphi_1$. When these have been solved, there
remains the explicit expression 
\begin{equation}
c=e^{-2\psi_0}2\varphi_1\psi_1
\end{equation}
for $\psi_1$ and the algebraic equation
\begin{equation}
f(\tau)=e^{-\psi_0}f'[\tau+\psi_0(\tau)]
\end{equation}
for $f'(\tau')$. After deleting the primes, we have
(\ref{standardapprox}). 

Dividing (\ref{1}) by (\ref{2}), assuming that $\varphi_1$ and
$\psi_0$ are periodic, and averaging, one obtains
\begin{equation}
\label{periodiccondition}
4q=\overline{\left({a\over b}\right)}.
\end{equation}
This determines the constant $q$. In order to solve (\ref{1}) and
(\ref{2}), note that they can be combined into a single first order
linear ODE:
\begin{equation}
\frac{d}{d\tau}(\varphi_1 e^{-2\psi_0}) - 
2\left(1+\frac{a}{4b}\right)\varphi_1 e^{-2\psi_0}
=-2(q+1)b .
\end{equation}
The equation $\dot y+fy=g$ with $f$ and $g$ periodic has a unique
periodic solution $y$ if and only if $\bar f\ne 0$
\cite{understanding}. In our case this condition is $q\ne
-1$. However, for $q=-1$ we have $g=0$, and so there still is a
solution that is unique up to an overall factor, $\ln y+\int f=\rm
const$. With $\varphi_1 e^{-2\psi_0}$ known, we can simply integrate
\begin{equation}
\dot\psi_0=b\left(\varphi_1 e^{-2\psi_0}\right)^{-1}-1
\end{equation}
to obtain $\psi_0$ up to a constant. The condition for $\psi_0$ to be
periodic is easily seen to be (\ref{periodiccondition}) again. Because
$\varphi_1 e^{-2\psi_0}$ is unique, the arbitrary additive constant in
$\psi_0$ corresponds to a constant factor in $\varphi_1$. Together
these changes leave (\ref{standardapprox}) invariant up to an overall
constant factor that can be thought of as a change of units (a change
of $l_0$).


\section{Curvature near the SSH}
\label{appendix:curvature}


We have found that near the SSH the spacetime metric can be
approximated by a metric of the form
\begin{equation}
ds^2= -du\,dv + r(u,v)^2 \,d\Omega^2.
\end{equation}
We define the null tetrad
\begin{eqnarray}
\nonumber
E_1&=&{\partial\over\partial u}, \qquad
E_2={\partial\over\partial v}, \\
\nonumber
E_3&=&{1\over r}{\partial\over\partial \theta}, \qquad
E_4={1\over r\sin\theta}{\partial\over\partial \varphi}. \\
\end{eqnarray}
It is parallely transported along any radial geodesic. This is obvious
because the spacetime is spherically symmetric and the reduced
spacetime is flat. The only non-vanishing tetrad components of the
Ricci tensor are
\begin{eqnarray}
R_{11}&=&-{2r_{,uu}\over r}, \\
R_{12}&=&R_{21}=-{2r_{,uv}\over r}, \\
R_{22}&=&-{2r_{,vv}\over r}, \\
R_{33}&=&R_{44}={1+4r_{,u}r_{,v}+4rr_{,uv}\over r^2}.
\end{eqnarray}
With
\begin{equation}
r(u,v)=f(puv^{q/p})^nv^{1/2p},
\end{equation}
where $f$, $q$, $p\equiv q+1$ and $n$ are constants, the non-vanishing
frame coefficients are
\begin{eqnarray}
R_{11}&=&-2n(n-1)u^{-2}, \\
R_{12}&=&-2n\left({1\over 2p}+n{q\over p}\right)u^{-1}v^{-1}, \\
R_{22}&=&-2\left({1\over 2p}+n{q\over p}\right)
\left({1\over 2p}+n{q\over p}-1\right)v^{-2}, \\
R_{33}&=&R_{44}=r^{-2} - 4 R_{12}.
\end{eqnarray}
The Ricci scalar is $R=-4R_{12}+2R_{33}$, and the Kretschmann scalar
is equal to $R^2$. We see that $R_{33}$ and $R$ always diverge at
$r=0$. For $n\ne 0$, $R_{11}$ or $R_{12}$ also diverge at
$u=0$. $R_{22}$ diverges at $v=0$, except for two specific values of
$q$. Note that one would miss the potential curvature singularities at
$u=0$ and $v=0$ if one only looked for the blowup of curvature
scalars.


\section{Brady Class I spacetimes}
\label{appendix:brady}


Starting from outgoing Bondi coordinates, Brady \cite{Brady} defines
the CSS coordinate $x=r/(-u)$. A natural choice of the log scale
coordinate is $\tau=-\ln(-u)$, and this gives
\begin{equation}
ds^2=e^{-2\tau}\left[-g(\bar g-2x)d\tau^2-2g\,dx\,d\tau+x^2\,d\Omega^2\right],
\end{equation}
where $g$ and $\bar g$ are functions of $x$ only.
Brady's class I solutions have the range $0\le x<\infty$. The
dynamical boundary $x=0$ is a regular center, where $g,\bar
g=1+O(x^2)$. As $x\to \infty$, $y\equiv \bar g/g\to 1/2$ and $z\equiv
x/\bar g\to 1/(1+c)$ where $c\equiv \sqrt{4\pi}\kappa>1$ is a
parameter of the solution. The metric as $x\to \infty$ becomes
\begin{equation}
ds^2\simeq e^{-2\tau}\left[-2(c^2-1)x^2d\tau^2
-4(1+c)x\,dx\,d\tau+x^2\,d\Omega^2\right].
\end{equation}
This can be compactified by $x\equiv \tilde x^{-n}$ for
any $n>0$, which gives
\begin{eqnarray}
\nonumber
ds^2 & \simeq & e^{-2\tau}\tilde x^{-2n-1}\Big[-2(c^2-1)\tilde x\,d\tau^2 \\
&+&4n(1+c)d\tilde x\,d\tau+\tilde x\,d\Omega^2\Big].
\end{eqnarray}
As $f_u\simeq (c-1)/(4n)\tilde x$, $\tilde
x=0$ is a fan, but the compactification shows that it is also a
dynamical null infinity.



\end{document}